\documentclass[12pt]{article}

\usepackage[a4paper,text={15cm,23cm},centering]{geometry}
\usepackage{latexsym}
\usepackage{amsmath}
\usepackage{amssymb}
\usepackage{graphicx}
\usepackage{mathrsfs}
\usepackage{eucal}
\usepackage{bm}
\usepackage{epsfig}
\usepackage{subcaption}
\usepackage{wrapfig}
\usepackage{color}

\usepackage[breaklinks=true,colorlinks=true,linkcolor=blue,urlcolor=blue,citecolor=blue]{hyperref}


\def\eps{\varepsilon}

\def\lth{\langle\!\langle}
\def\rth{\rangle\!\rangle}

\def\Zh{\widehat{Z}}

\def\D{{\rm D}}
\def\lth{\langle\!\langle}
\def\rth{\rangle\!\rangle}

\usepackage{titletoc}
\begin{document}

\title{Dispersive Dynamics in the Characteristic Moving Frame}
\author{D.J. Ratliff}



\maketitle
\begin{abstract}
A mechanism for dispersion to automatically arise from the dispersionless Whitham Modulation equations (WMEs) is presented, relying on the use of a moving frame. The speed of this is chosen to be one of the characteristics which emerge from the linearisation of the Whitham system, and assuming these are real (and thus the WMEs are hyperbolic) morphs the WMEs into the Korteweg - de Vries (KdV) equation in the boosted co-ordinate. Strikingly, the coefficients of the KdV equation are universal, in the sense that they are determined by abstract properties of the original Lagrangian density. Two illustrative examples of the theory are given to illustrate how the KdV may be constructed in practice. The first being a revisitation of the derivation of the KdV equation from shallow water flows, to highlight how the theory of this paper fits into the existing literature. The second is a complex Klein-Gordon system, providing a case where the KdV equation may only arise with a moving frame.
\end{abstract}

\section{Introduction}

The Whitham Modulation equations (WMEs) are a set of quasilinear first order partial differential equations (PDEs) which emerge as a tool in the analysis of nonlinear waves across various fields of mathematical and physical interest~\cite{w65,wlnlw}. The approach focusses on (but is not restricted to) problems generated by a Lagrangian,
\begin{equation}\label{LagBasic}
\mathscr{L}(U) =  \iint_\Gamma \mathcal{L}(U,U_x,U_t,\ldots)\,dx\,dt\,,
\end{equation}
for state vector $U(x,t)$, $\Gamma$ denotes the box $[x_1,x_2]\times[t_1,t_2]$ on which the problem is posed and the lower dots signify that the Lagrangian density $\mathcal{L}$ may also depend on further derivatives of $U$. The assumption is that there exists a periodic wavetrain which minimizes the above Lagrangian of the form
\[
U = \hat{U}(kx+\omega t+\theta_0;k,\omega) \equiv \hat{U}(\theta;k,\omega)\,, \quad \hat{U}(\theta+2 \pi) = \hat{U}(\theta)\,,
\]
for phase $\theta$, wavenumber $k$ and frequency $\omega$.
The Whitham methodology then treats the phase, wavenumber and frequency as slowly varying, so that
\[
\theta = \eps^{-1}\Theta(X,T)\,, \quad k(X,T) = \Theta_X\,, \quad \omega(X,T) = \Theta_T\,,
\]
for slow variables $X = \eps x$ and $T = \eps t$ with $\eps \ll 1$ a small parameter.
From this substitution, one derives equations governing their evolution, typically by considering variations of the Lagrangian density averaged over a period of the wavetrain.
This leads to the nonlinear set of PDEs for the evolution of the wavenumber and frequency,
\begin{equation}\label{WME-Intro}
\mathscr{A}\big(k,\omega\big)_T+\mathscr{B}\big(k,\omega\big)_X = 0\,, \quad K_T = \Omega_X\,,
\end{equation}
The first equation represents the conservation of wave action for the original wave, with the functions $\mathscr{A}$ and $\mathscr{B}$ being the wave action and wave action flux of the original problem evaluated for the wavetrain $\hat{U}$ respectively. The second equation represents the conservation of waves, emerging from the necessity of cross derivatives of the phase $\Theta$ commuting.
With these equations, a wide variety of problems can be investigated, such as dispersive shock waves \cite{egs06,eh15,hac06}, the evolution of wavepackets \cite{k90,wj17} and modulational stability \cite{bjn10,bhj16,kp00}, demonstrating both their versatility and utility. 

The modulational stability properties of the original wave can be found based on the classification of the WMEs (\ref{WME-Intro}). The primary diagnostic for this is the eigenvalues which emerge from the linearisation of (\ref{WME-Intro}) about a fixed wavenumber and frequency, which are denoted as the \emph{characteristics} of the WMEs. If these are purely real, then the WMEs are hyperbolic and the original wavetrain is stable, whereas the presence of complex characteristics signify that the WMEs are elliptic and the wavetrain is in fact modulationally unstable. What distinguishes these two regimes is the sign of the \emph{Lighthill determinant}~\cite{l65,l67},
\begin{equation}\label{Lighthill_Det_Intro}
\Delta_L(k,\omega) := \mathscr{B}_k\mathscr{A}_\omega-\mathscr{A}_k\mathscr{B}_\omega\,,
\end{equation}
which is negative in the hyperbolic regime and positive when the WMEs are elliptic.
The link between the classification of the WMEs and stability was initially noted by Whitham himself, when considering the experimental observations of Benjamin and Feir, where ``the penny dropped'' regarding the connection between stability and classification of (\ref{WME-Intro})~\cite{w67,ms16}. The main purpose of this paper is to extend the use of these characteristics beyond stability, and show that these actually infer more about the wavetrain than its stability. Namely, we show that the characteristics have a fundamental role in the way in which the phase dynamics evolve, as well as how these lead to dispersive dynamics automatically emerging within the WMEs.

Within the modulation of wavetrains, the emergence of dispersive effects from the WMEs has been a persisting problem. Such phenomena is required to resolve various singularities which may occur, such as gradient catastrophes (for example, see \cite{egl12,eh15,w65}). Conventionally, the justification to include these effects is to utilise the dispersion relation to determine the dispersive term one should add, without a formal derivation \cite[\S 15.2]{wlnlw}. Some advances towards rigourously deriving the dispersive corrections to the Whitham system have been made, showing how one may morph the Whitham modulation equations into dispersive equations such as the Korteweg-de Vries (KdV) equation and other long wave models \cite{b13,rb15,r17,bspmnw}, which require the linearisation of the system to be degenerate in some way (such as the appearance of zero characteristics). The criteria for these conditions, henceforth referred to as criticality, can often be formulated using the conservation law components $\mathscr{A}$ and $\mathscr{B}$. These may then be satisfied by a suitable choice of fixed wavenumber $k_0$ and frequency $\omega_0$.

The altered approach proceeds in the following way. Given the same class of problem, again with wavetrain solution $\hat{U}$,
the modified modulation procedure utilises the ansatz
\begin{equation}\label{bridges-ansatz}
U = \hat{U}\big(\theta+\eps \phi(X,T);k_0+\eps^2 q(X,T),\omega_0+\eps^3 s(X,T)\big)+\eps^3 W(\theta,X,T)\,,
\end{equation}
for $X = \eps x,\,T = \eps^3 t$ and $\eps \ll 1$. The slowly varying phase perturbation $\phi$ is analogous to the slow phase from Whitham theory, albeit with different scales, and is related to the functions $q$ and $s$ via
\[
q = \phi_X\,, \quad s = \phi_T\,, \quad \Rightarrow \quad q_T = s_X\,.
\]
The reduction procedure is then undertaken in a standard way, where the above ansatz is substituted into the governing equations, one expands about the $\eps = 0$ point and solves the resulting problem at each order of $\eps$. The advantage of the above form (\ref{bridges-ansatz}) is that it introduces derivatives of the wavetrain into the analysis, leading to several simplifications throughout. Utilising this approach leads to the KdV equation
\begin{equation}\label{kdv-bridges}
2\mathscr{A}_k(k_0,\omega_0)\,q_T+\mathscr{B}_{kk}(k_0,\omega_0)\,qq_X+ \mathscr{K}(k_0,\omega_0)\,q_{XXX}=0\,,
\end{equation}
requiring that the condition that the wavenumber and frequency are chosen so that
\[
\mathscr{B}_k(k_0,\omega_0) = 0\,.
\]
An immediate, striking feature of the above KdV is that the majority of its coefficients are derivatives of the conservation law components, owing to the proposed form of the ansatz (\ref{bridges-ansatz}). The last coefficient emerges from a linear algebraic analysis~\cite{bspmnw}. The criteria that leads to the KdV also corresponds to the Whitham system having a zero characteristic. Thus, the KdV equation emerges in universal form in the same sense as the Whitham equations, namely that the coefficients are tied to abstract properties of the Lagrangian which generates the system.

Often though, there are ways to construct the reduction procedure in order to avoid having to impose such conditions on the wavenumber and frequency of the original wave, and this is typically done instead by a careful choice of travelling co-ordinates. The idea of using a moving reference frame to subvert criticality is not unusual. In the context of water waves, this is frequently used in the derivation of the KdV there through choosing the gravity wave speed~\cite{d97,g05,m81}. The novelty here is to demonstrate that the same idea, that is by the use of a suitable wavespeed, one may automatically cause dispersion to enter the phase dynamics. It transpires that the moving frame required in order to introduce dispersion into the modulation theory is exactly one of the characteristics admitted by (\ref{WME-Intro}). The only restriction that emerges, therefore, is that these speeds be real on physical grounds. This essentially imposes that the system (\ref{WME-Intro}) is hyperbolic local to the point $(k_0,\omega_0)$, and no further restrictions are required.

The approach to do so is a natural alteration of the ansatz (\ref{bridges-ansatz}) to the moving frame. This travelling coordinate requires one to slightly change how the frequency of the wave are perturbed, in line with the phase consistency condition. Overall, the new ansatz this paper adopts is
\begin{equation}\label{new-ansatz}
U = \hat{U}\big(\theta+\eps \phi(X,T);k_0+\eps^2 q(X,T),\omega_0+\eps^2 c q(X,T)+\eps^3 s(X,T)\big)+\eps^3 W(\theta,X,T)\,,
\end{equation}
where instead $X = \eps(x+ct)$ and $T = \eps^3t$, and $c$ is the speed of the moving frame. The functions $q$ and $s$ remain as defined before. The main result of the paper will be to show that, providing $c$ is a real characteristic of the WMEs, the modulation approach admits precisely the KdV equation
\begin{equation}
2 \sqrt{-\Delta_L(k_0,\omega_0)}q_T+(\partial_k+c\partial_\omega)^2(\mathscr{B}+c\mathscr{A})(k_0,\omega_0) qq_X+\mathscr{K}(k_0,\omega_0)q_{XXX} = 0\,.
\end{equation}
The coefficients once again relate to abstract properties of the Lagrangian - the first due to the definition (\ref{Lighthill_Det_Intro}), the second is clear and the third emerges again from a linear analysis. This nce again makes the emergent KdV universal. Moreover, the criterion that must be met in order for the above KdV to emerge is weaker that that of (\ref{kdv-bridges}), meaning that one expects the KdV in the moving frame to be more prevalent and applicable within applications.

This paper offers two insightful applications of the methodology. The first is a revisitation of the emergence of the KdV equation from shallow water hydrodynamics in a moving frame, so that this derivation may be reformulated using the theory presented in the paper. This is so that a connection to the existing literature can be made and the approach presented here better understood. The second will be to use the cubic Complex Klein Gordon (CKG) equation, which will illustrate how the theory provides even further instances of the emergence of dispersion from the modulation theory. Prior to this paper, the existing literature cannot show the presence of KdV dynamics from this system, and so the theory presented here leads to further instances of the KdV as an applicable model in nonlinear wave problems.

An outline of the paper is as follows. In \S\ref{sec-WMT} the relevant Whitham Modulation Theory is reviewed, including a discussion of characteristics and their nature. Subsequently, in \S\ref{sec-mod}, the details of the modulation approach which leads to dispersion emerging in the moving frame are presented. Two examples of how the theory may be applied are  then presented. In \S\ref{sec-swh}, we revisit the well-known emergence of the KdV equation in shallow water hydrodynamics in order to demonstrate how the result of this paper fits into the existing literature. The second example \S\ref{sec-ckg}, demonstrating how the results of this paper extend the applicability of the KdV equation to additional nonlinear wave problems. Concluding remarks are presented in \S\ref{sec-cr}

\section{Whitham Averaging, Modulation and Local Hyperbolicity}\label{sec-WMT}
The starting point for the required theory of this paper will be the class of problems generated by a Lagrangian density (\ref{LagBasic}).
In the later reduction procedure, it is more convenient to instead deal with the \emph{multisymplectic} version of this Lagrangian:
\begin{equation}\label{LagMSF}
\mathscr{L}(Z) = \iint_\Gamma \frac{1}{2}\big \langle {\bf M}Z_t,Z\rangle+\frac{1}{2}\langle {\bf J}Z_x,Z\rangle-S(Z)\, dx\,dt\,,
\end{equation}
for new state vector $Z(x,t)$. This form is obtained via a sequence of Legendre transformations involving terms containing space and time derivatives of the state vector. By doing this, the skew symmetric matrices ${\bf M}$ and ${\bf J}$ enter into the problem and the symplectic structure of the Lagrangian is encoded within them \cite{b97}. The remaining part of the Lagrangian density, $S$, may be thought of as a Hamiltonian function which is independent of derivatives of $Z$. It is this casting of the Lagrangian which ultimately leads to the connection between the conservation law components and the resulting coefficients arising from the reduction procedure, however we note that this form isn't required for the discussion on the Whitham averaging procedure below.

The Whitham averaging principle proceeds as follows. We start by taking a solution of the Euler-Lagrange equations associated with (\ref{LagMSF}) of the form
\[
Z(x,t) = \Zh(kx+\omega t+\theta_0;k,\omega) \equiv \Zh(\theta;k,\omega)
\]
for constants $k,\,\omega,\, \theta_0$. Solutions of this form are usually periodic travelling waves, however there is a much more general class of solutions parameterised by variables like $\theta$ known as \emph{relative equilibria} \cite{bspmnw}. Such solutions are associated to a continuous symmetry of the Lagrangian, such as invariance under affine actions or rotations. The next step towards obtaining the Whitham modulation equations is to substitute this solution into the Lagrangian (\ref{LagMSF}) and average over the phase $\theta$, resulting in
\begin{equation}\label{AvgLag}
\widehat{\mathscr{L}}(k,\omega) = \frac{1}{2 \pi}\int_0^{2\pi}\frac{\omega}{2}\langle {\bf M}\Zh_\theta,\Zh\rangle +\frac{k}{2}\langle {\bf J}\Zh_\theta,\Zh\rangle - S(\Zh) \, d \theta\,.
\end{equation}
The final step is to consider the case where the phase $\theta$ is instead a rapid phase which slowly varies in both space and time, so that $\theta = \eps^{-1}\Theta(X,T)$, with $X = \eps x,\,T = \eps t$ and $\eps \ll 1$. This also naturally leads to the definition of the slowly varying wavenumbers and frequencies
\[
K(X,T) = \Theta_X\,, \quad \Omega(X,T) = \Theta_T\,, \quad \implies K_T = \Omega_X\,.
\]
By substituting these into the averaged Lagrangian (\ref{AvgLag}) and taking variations with respect to $\Theta$, one arrives at the \emph {Whitham Modulation Equations} (WMEs)
\begin{equation}\label{WME}
\mathscr{A}(K,\Omega)_T+\mathscr{B}(K,\Omega)_X = 0\,, \quad \mbox{with} \quad K_T = \Omega_X\,.
\end{equation}
with
\begin{equation}\label{Cons-Laws}
\mathscr{A} = \mathscr{L}_\omega = \frac{1}{4 \pi}\int_0^{2 \pi}\langle \Zh, {\bf M}\Zh_\theta\rangle \equiv \frac{1}{2} \lth \Zh,{\bf M}\Zh_\theta\rth\,, \quad \mathscr{B} =\mathscr{L}_k =  \frac{1}{2}\lth \Zh,{\bf J}\Zh_\theta\rth\,.
\end{equation}

The stability properties of the constant wavenumber and frequency state $(k_0,\omega_0)$ of the Whitham modulation equations (\ref{WME}) may be assessed via linearisation. Indeed, setting $K = k_0+\delta q(X,T),\,\Omega = \omega_0+\delta s(X,T)$ with $\delta \ll 1$ and only retaining $\mathcal{O}(\delta)$ terms. The resulting linear system may be written in the form:
\[
\begin{pmatrix}
q\\
s
\end{pmatrix}_T+\frac{1}{\mathscr{A}_\omega}
\begin{pmatrix}
0&-\mathscr{A}_\omega\\
\mathscr{B}_k&\mathscr{A}_k+\mathscr{B}_\omega
\end{pmatrix}
\begin{pmatrix}
q\\
s
\end{pmatrix}_X = {\bf 0}\,,
\]
where the conservation law components are evaluated at $(k_0,\omega_0)$ and we assume that $\mathscr{A}_\omega \neq 0$. The typical normal mode ansatz $(q,s)^T = (q_0,s_0)^Te^{i(X+c T)}$ used to determine the stability of the problem, and result in the eigenvalue $c$ (also referred to as the system's \emph{characteristic speeds}) satisfying the quadratic
\begin{equation}\label{c-defn}
\mathscr{A}_\omega c^2+(\mathscr{A}_k+\mathscr{B}_\omega)c+\mathscr{B}_k = 0\,,
\end{equation}
and thus, by using (\ref{Cons-Laws}), are given by
\begin{equation}\label{c-value}
c = \frac{1}{\mathscr{A}_\omega}\big[-(\mathscr{A}_k+\mathscr{B}_\omega) \pm \sqrt{-\Delta_L}\big]\,, 
\end{equation}
where we have introduced the Lighthill determinant
\[
\Delta_L \equiv \mathscr{A}_\omega\mathscr{B}_k-\mathscr{A}_k\mathscr{B}_\omega\,.
\]
The Lighthill determinant is used primarily as a diagnostic for the nature of the Whitham equations locally~\cite{l67}, and will appear within the subsequent theory of this paper. For $\Delta_L>0$ both characteristic speeds are complex and the original wave is unstable, and the Whitham modulation equations are said to be \emph{locally elliptic}. Conversely, whenever $\Delta_L<0$ these speeds are real, the wavetrain is stable and the Whitham modulation equations are referred to as \emph{locally hyperbolic}. This latter regime is the one of interest within this paper, as we will show that so long as the basic state being considered has the property of local hyperbolicity then it follows that there are two associated KdV equations which emerge, one for each characteristic speed appearing in (\ref{c-value}).

\section{Modulation Reduction}\label{sec-mod}
With the knowledge of the characteristics, we are in a position to construct a more informed ansatz for the modulation theory. This is done in such a way that causes the condition (\ref{c-defn}) to arise naturally as a solvability condition during the reduction procedure.
To this end, we utilise the ansatz
\begin{equation}\label{ansatz}
Z = \Zh\big(\theta+\eps \phi(X,T),k+\eps^2q(X,T),\omega+\eps^2cq+\eps^4\Omega(X,T)\big)+\eps^3W(\theta,X,T)\,,
\end{equation}
for the slow variables
\[
X = \eps(x+ct)\,, \quad T = \eps^3 t\,.
\]
The wavespeed $c$ is assumed to satisfy (\ref{c-defn}), which will be highlighted during the reduction procedure. The modulation functions $\phi,\,q$ and $s$ are related via
\[
\phi_X = q\,, \quad \phi_T = s\,, \quad \Rightarrow \quad q_T= s_X\,.
\]

The methodology is to substitute the above ansatz into the multisymplectic Euler Lagrange equation
\[
{\bf M}Z_t+{\bf J}Z_x = \nabla S(Z)\,,
\]
and Taylor expand about the $\eps = 0$ point. The above Euler-Lagrange equations, with the multisymplectic structure, lead to several simplifications within the analysis. We briefly recount the ones necessary for this paper, but the full details of these may be found in \cite{br17}. By substituting the wavetrain solution into the above, one obtains the ODE
\[
(\omega{\bf M}+k{\bf J})\Zh_\theta = \nabla S(\Zh).
\]
By differentiating this with respect to $\theta,\,k$ and $\omega$, one obtains the following results:
\begin{equation}\label{First_Derivs}
{\bf L}\Zh_\theta = {\bf 0}\,, \quad {\bf L}\Zh_k = {\bf J}\Zh_\theta\,, \quad {\bf L}\Zh_\omega = {\bf M}\Zh_\theta\,,
\end{equation}
where we have introduced the linear operator
\[
{\bf L}:=\D^2S(\Zh)-(\omega {\bf M}+k{\bf J})\partial_\theta\,.
\]
The first result identifies that $\Zh_\theta \in \ker({\bf L})$, which within this analysis is assumed to be no larger. A consequence of this is that for inhomogeneous problems,
\[
{\bf L}F = G \quad \mbox{is solvable if and only if} \quad \lth \Zh_\theta,G\rth = 0\,.
\] 
The remaining two will be used to simplify the analysis. One can identify these as the start of a Jordan chain, each involving the skew symmetric matrices ${\bf M}$ and ${\bf J}$, however the theory of this is not necessary within this paper. For further details, the author directs the reader to~\cite{bspmnw}.

 The result of this approach will be the emergence of the KdV equation without the need to impose specific constraints on the conservation law components, as is the case with other modulational analyses\cite{b13,b14,rb15}. Instead, all that is required is the local hyperbolicity of the linearised Whitham equations so that the wavespeed is real, and the KdV equation automatically follows in the moving frame. Below the details of the modulation reduction are provided, demonstrating the key steps which lead to the emergence of the KdV equation.

\subsection{Leading to second order}
Leading order is just the defining equation for $\Zh$ as expected, and the first order in $\eps$ is
\[
\phi {\bf L}\Zh_\theta = 0\,.
\]
This is always true for $\Zh$ by (\ref{First_Derivs}). The following order in $\eps$ also gives
\[
q{\bf L}(\Zh_k+c{\bf M}\Zh_\omega) = ({\bf J}+c{\bf M})\Zh_\theta\,,
\]
which is also true from properties of the basic state outlined in (\ref{First_Derivs}).

\subsection{Third Order}
The remaining terms at third order give
\[
{\bf L}W_0 = q_X({\bf J}+c{\bf M})(\Zh_k+c\Zh_\omega)\,,
\]
where we have expanded $W$ as
\[
W = \sum_{n=0}^\infty\eps^nW_n(\theta,X,T)\,.
\]
Solvability gives
\[
\lth \Zh_\theta,{\bf J}(\Zh_k+c\Zh_\omega)+c{\bf M}(\Zh_k+c\Zh_\omega)\rth = \mathscr{B}_k+(\mathscr{A}_k+\mathscr{B}_\omega)c+\mathscr{A}_\omega c^2 =0\,.
\]
This is true by the choice of $c$ from (\ref{c-defn}), and so we can continue the analysis. We may then define $\xi_3$ with
\[
{\bf L}\xi_3 = ({\bf J}+c{\bf M})(\Zh_k+c\Zh_\omega)\,.
\]
Interestingly, $\xi_3$ may be thought of as the third term of a twisted Jordan chain, the termination of which leads to the emergence of dispersion of the problem. A discussion of this is not required to obtain the KdV equation sought here, and the reader is instead directed to \cite{br17} for details and implications of this connection. 
Overall, we may then write that
\[
W = q_X \xi_3\,.
\]

\subsection{Fourth Order}
At fourth order, we have that
\[
{\bf L}(W_1-\phi q_X(\xi_3)_\theta) = \Omega{\bf M}\Zh_\theta-\phi_T{\bf L}\Zh_\omega+q_{XX}({\bf J}+c{\bf M})\xi_3\,.
\]
The first two terms cancel each other out, and the last is solvable because the zero eigenvalue of ${\bf L}$ is even~\cite{am78}. Thus the system has the solution
\[
W_1 = \phi q_X(\xi_3)_\theta+q_{XX}\xi_4\,, \quad {\bf L}\xi_4=({\bf J}+c{\bf M})\xi_3\,.
\]

\subsection{Fifth Order}
At the final order, one has the equation
\[
\begin{split}
{\bf L}\widetilde{W}_2 =& q_T({\bf M}\Zh_k+c{\bf M}\Zh_\omega)+\Omega_X\big({\bf J}\Zh_\omega+c{\bf M}\Zh_\omega)\\
&qq_X\bigg(({\bf J}+c{\bf M})\big( \Zh_{kk} + 2c \Zh_{\omega k} + c^2\Zh_{\omega\omega}\big)-\D^2S(\Zh)(\Zh_k+c\Zh_\omega,\xi_3) \bigg)\\
&+q_{XXX}({\bf J}+c{\bf M})\xi_4\,.
\end{split}
\]
The notation $\widetilde{W}_2$ denotes $W_2$ along with a collection of all the terms which can be shown to lie in the range of ${\bf L}$ at this order, and its exact form isn't important to the analysis. Appealing to solvability at this stage, the dispersive term $q_{XXX}$ has the coefficient
\[
\lth \Zh_\theta,({\bf J}+c{\bf M})\xi_4\rth \equiv -\mathscr{K}\,,
\]
by definition, and arising as the coefficient of the nonlinear term one has
\[
\lth \Zh_\theta,({\bf J}+c{\bf M})\big( \Zh_{kk} + 2c \Zh_{\omega k} + c^2\Zh_{\omega\omega}\big)-\D^2S(\Zh)(\Zh_k+c\Zh_\omega,\xi_3) \rth = -(\partial_k+c\partial_\omega)^2(\mathscr{B}+c\mathscr{A})\,,
\]
with $c$ fixed.
Finally, the coefficient which arises from the time term, upon using $q_T = \Omega_X$ gives
\[
\lth \Zh_\theta,{\bf M}\Zh_k+c{\bf M}\Zh_\omega+{\bf J}\Zh_\omega+c{\bf M}\Zh_\omega\rth = \mathscr{A}_k+\mathscr{B}_\omega+2c\mathscr{A}_\omega = \pm 2\sqrt{\mathscr{A}_k\mathscr{B}_\omega-\mathscr{A}_\omega\mathscr{B}_k} \equiv \pm 2\sqrt{-\Delta_L}\,,
\]
which is seen by rearranging (\ref{c-value}). Overall, these results combine to give that the fifth order analysis is solvable precisely when $q$ satisfies the KdV equation
\begin{equation}\label{KdV-uni}
\pm 2\sqrt{-\Delta_L}q_T+(\partial_k+c\partial_\omega)^2(\mathscr{B}+c\mathscr{A})qq_X+\mathscr{K}q_{XXX} = 0\,.
\end{equation}

Thus, to summarise, the KdV equation (\ref{KdV-uni}) emerges from the ansatz (\ref{ansatz}) precisely when $c$ is a real characteristic of the WMEs (\ref{WME}), with the only additional impositions being that all of its coefficients are nonzero. This therefore suggests a more prevalent emergence for the KdV from the modulation approach, as the conditions required to obtain it are more relaxed compared to previous analyses. For example, in the work of Bridges~\cite{b13} the KdV equation was shown to arise providing the condition
\[
\mathscr{B}_k(k_0,\omega_0) = 0\,,
\]
is satisfied, corresponding to the WMEs possessing a zero characteristic. However, it is not always the case that a zero characteristic emerges from a given WMEs, with an example of such a case presented in \S\ref{sec-ckg}, and so the theory of this paper extends the scenarios for which a KdV is obtainable.

\section{Example 1 - Shallow Water Hydrodynamics}\label{sec-swh}
In order to best illustrate how the theory of this paper fits in with the current literature, we revisit a very well known example where it exists, albeit in an unabstracted format. This occurs within the shallow water system, given by
\begin{equation}\label{SW}
\begin{split}
\eta_t+(\eta u)_x = 0\,,\\
u_t+uu_x+g \eta_x +\frac{g \eta_0^2}{3}\eta_{xxx} = 0\,,
\end{split}
\end{equation}
for horizontal fluid velocity $u(x,t)$, free surface elevation $\eta(x,t)$, gravity $g$ and quiescent fluid height $\eta_0$. In particular, we concern ourselves with the case of irrotational flow, so that we may introduce a velocity potential $\phi(x,t)$ with $u = \phi_x$, which allows one to integrate the second equation of (\ref{SW}) with respect to $x$. This results in the potential shallow water system
\begin{equation}\label{SWphi}
\begin{split}
\eta_t+(\eta \phi_x)_x = 0\,,\\
\phi_t+\frac{1}{2}\phi_x^2+g \eta +\frac{g \eta_0^2}{3}\eta_{xx} = R\,,
\end{split}
\end{equation}
for some constant $R$, which may be thought of as the total head of the flow.

The emergence of the KdV from this system (\ref{SW}) is well documented (\cite{m81,d11,g05} and references therein), and is achieved by using the multiple scales expansion
\[
\begin{array}{rl}
u = \eps^2 U(X,T)+\eps^4 V(X,T)\,, & \eta = \eta_0+ \eps^2 H(X,T)+\eps^4 G(X,T)\,, \\[3mm]
 X = \eps(x\pm \sqrt{g \eta_0}t)\,, & T = \eps^3t\,.
\end{array}
\]
The asymptotic analysis then gives both that $H = \mp\sqrt{\frac{\eta_0}{g}}U$ and that $U$ must satisfy the KdV equation
\[
\pm 2\sqrt{g \eta_0}U_T\pm 3\sqrt{g\eta_0}UU_X+\frac{g \eta_0^3}{6}U_{XXX} = 0\,.
\]
The aim will be to demonstrate how the KdV emerges from the perspective of this paper, recovering the above result. We do so by modulating the relative equilibrium associated with the affine symmetry of the velocity potential, due to the fact that one is free to add a constant to $\phi$ and leave the solution unaltered. The uniform solution may therefore be written as
\[
\phi = kx+\omega t+\theta_0 \equiv \theta\,, \quad \implies \quad u = k\,, \quad \eta = \eta_0 = g^{-1}\bigg(R-\omega-\frac{k^2}{2}\bigg)\,.
\]
The conservation law associated with this symmetry is the conservation of mass, which is simply the first equation of (\ref{SWphi}), which gives the conservation law components as
\[
\mathscr{A}(k,\omega) = \eta_0\,, \quad \mathscr{B}(k,\omega) = k \eta_0\,.
\]
Following the theory of this paper, we compute the relevant wavespeed for the emergence of the KdV. This must satisfy the quadratic
\[
-g^{-1}c^2-\frac{2k}{g}c+g^{-1}(g\eta_0-k^2) = 0\,,
\]
which gives the wavespeeds as
\[
c = -k \pm \sqrt{g \eta_0}\,.
\]
This is exactly the moving frame required for the KdV to emerge in the literature, and so this result is unsurprising, but it highlights how the framework given earlier fits within the existing theory.

All that remains is to determine the coefficients of the KdV according to the theory of the present paper. The coefficient of the time derivative term is simply
\[
\pm 2\sqrt{-\Delta_L} = \pm \frac{2}{g}\sqrt{g \eta_0}\,.
\] 
The coefficient of the nonlinear term may be computed as
\[
(\partial_k+c\partial_\omega)^2(\mathscr{B}+c\mathscr{A}) = \pm\frac{3}{g}\sqrt{g \eta_0}\,.
\]
Finally, the dispersive term may be computed using the relevant Jordan chain theory. This is achieved by abridging the analysis of Bridges~\cite{b13} and gives the coefficient
\[
\mathscr{K} = \frac{\eta_0^3}{3}\,.
\]
Thus, we obtain the KdV equation for the velocity
\begin{equation}
\pm \sqrt{\frac{\eta_0}{g}}\bigg(2q_T+3qq_X\bigg)+\frac{\eta_0^3}{3}q_{XXX} = 0\,,
\end{equation}
The free surface version may be obtained by using the transformation $H = \pm \sqrt{\frac{\eta_0}{g}}\,q$, giving
\begin{equation}
H_T\pm\frac{1}{2}\sqrt{\frac{g}{\eta_0}}\bigg(\frac{3}{2}H^2+\frac{\eta_0^3}{3}H_{XX}\bigg)_X = 0\,,
\end{equation}
which is exactly the KdV equation obtained by Korteweg and De Vries, as well as various others since \cite{m81,d11}. Thus, the theory of this paper is consistent with the known liter

\section{Example 2 - Complex Klein Gordon Equation}\label{sec-ckg}
The second, and most illustrative, application of the theory will be the cubic complex Klein Gordon (cCKG) equation. This is given by
\begin{equation}\label{CKG}
\alpha\Psi_{tt}+\Psi_{xx}-\Psi+ |\Psi|^2\Psi = 0\,,
\end{equation}
for complex-valued function $\Psi(x,t)$ and real constant $\alpha$. This equation, for $\alpha = -1$, emerges in scenarios where subharmonic instabilities arise, with the most notable example being the Kelvin-Helmholtz instability \cite{bb97,kl95,w79}. The principle aim will be to show that the use of the moving frame will be the \emph{only} way to obtain the KdV equation for this system for nontrivial relative equilibria, owing to the fact that this system is second order in time.

The solution from which the KdV equation will emerge is that which is associated with the $SO(2)$ symmetry of (\ref{CKG}), which is the plane wave solution
\[
A = \Psi_0e^{i \theta}\,, \quad {\rm with} \quad |\Psi_0|^2 = 1+\alpha\omega^2+k^2\,.
\]
In order to apply the theory of this paper, we need to consider the conservation law associated with (\ref{CKG}), which may be found as
\begin{equation}
\mathscr{A}(k,\omega) = \alpha\omega|\Psi_0|^2 = \alpha\omega(1+\alpha\omega^2+k^2)\,, \quad \mathscr{B}(k,\omega) = k|\Psi_0|^2  = k(1+\alpha\omega^2+k^2)\,.
\end{equation}
At this stage, it can be made clear that the moving frame is necessary in order to reduce the system to a KdV equation about this wave. Utilising the theory of \textsc{Bridges}~\cite{b13}, the KdV equation emerges at points in $(k,\omega)$-space where
\[
\mathscr{B}_k = |\Psi_0|^2+2k^2 = 0\,.
\]
This cannot be satisfied for any choice of the parameter values, since the amplitude of the wave needs to be positive. Thus, the KdV equation cannot emerge in the fixed frame.

The required derivative to compute the wavespeed $c$ are given by
\[
\mathscr{A}_\omega = \alpha|\Psi_0|^2+2\alpha^2\omega^2\,, \quad \mathscr{A}_k = \mathscr{B}_\omega = 2\alpha\omega k\,, \quad \mathscr{B}_k =|\Psi_0|^2+2k^2\,.
\]
These can then be used to construct the polynomial which defines $c$, given in (\ref{c-defn}), resulting in the wavespeeds
\[
c_\pm = \frac{-2 \alpha \omega k\pm \sqrt{\alpha|\Psi_0|^2(2-3 |\Psi_0|^2)}}{ \alpha|\Psi_0|^2+2\alpha^2\omega^2}\,,
\]
which are real when $\alpha<0\,, \ |\Psi_0|^2>\frac{2}{3}$.  

In this case, the theory of the paper may be used to construct the relevant KdV equation. The coefficient of the time derivative term is simply given by
\[
\pm2\sqrt{-\Delta_L} = \pm2\sqrt{\alpha|\Psi_0|^2(2-3 |\Psi_0|^2)}\,.
\]
There are two possibilities for the quadratic term, one for each sign in the wavespeed, and gives
\[
\begin{split}
(\partial_k+c_\pm\partial_\omega)^2(\mathscr{B}+c_\pm \mathscr{A}) =& \ 6(1+\alpha c_\pm ^2)(k+\alpha c_\pm \omega)\,.\\
\end{split}
\]
Finally, the dispersive coefficient may be computed using the Jordan chain approach as detailed in Bridges and Ratliff~\cite{br17}, giving that
\[
\begin{split}
\mathscr{K}_\pm =& \frac{(1+\alpha c^2_\pm)(k+\alpha c_\pm\omega)^2}{|\Psi_0|^2}\,.\\
\end{split}
\]
This fully determines the relevant KdV equations for each wavespeed, given by
\[
\pm2\sqrt{\alpha|\Psi_0|^2(2-3 |\Psi_0|^2)}\, q_T +6(1+\alpha c_\pm ^2)(k+\alpha c_\pm\omega)\,qq_X +\frac{(1+\alpha c_\pm^2)(k+\alpha c_\pm\omega)^2}{|\Psi_0|^2}\,q_{XXX} = 0\,.
\]
\section{Concluding Remarks}\label{sec-cr}
This paper has demonstrated that dispersion naturally emerges from the initially dispersionless Whitham modulation theory with no constraints other than hyperbolicity, providing one chooses a moving frame with a speed equal to one of the characteristics of the Whitham equations. This extends the range of scenarios in which the KdV arises, subverting the need for the wave parameters to satisfy specific conditions which would primarily be a vanishing characteristic, making its emergence more generic. Moreover, the theory of this paper highlights that the moving frame allows one to obtain a KdV equation in the moving frame for systems which are unable to support them in the fixed frame, suggesting that the KdV equation is perhaps more prevalent that previously thought.

The result of this paper, and its extensions, highlight that properties of the characteristics have a significant role in the nonlinear phase dynamics emerging from the modulation. For example, the multiplicities of the roots of (\ref{c-defn}) have an important role over the temporal properties emergent dynamics. When each root is distinct, as in the case presented here, the dynamics are unidirectional. This changes at points where two characteristics coalesce, where the bidirectional two-way Boussinesq equation becomes operational \cite{br17}. Further, we also have the possibility that the coefficient of the nonlinear of dispersive terms of the KdV (\ref{KdV-uni}) may vanish when the characteristics take specific values. In such cases, the ansatz must be rescaled and instead other nonlinear dispersive PDEs are expected to emerge from the reduction, such as the modified KdV or the fifth order KdV equations.

There are several natural extensions to the results covered here. Most readily, it may be extended to additional spatial dimensions by utilising the theory appearing within other works \cite{br18}, expecting to lead to a Kadomstev-Petviashvili equation in the moving frame. More interestingly, with the theory determined in the single phase case, the multiple phase analogue is relatively easy to generate. In fact, early work on the modulation approach with a moving frame in this setting has been developed for the case of coalescing characteristics \cite{br19}, and the more generic case of hyperbolicity is expected to be simpler and more pervasive. 

\section*{Acknowledgements}
The author would like to thank Prof. Tom Brides and Prof. Gennady El for their invaluable discussions throughout the formulation of this paper.

\end{document}